\documentclass[12pt]{iopart}

\usepackage{graphicx}
\begin{document}
\def\be{\begin{equation}}
\def\ee{\end{equation}}

\def\bc{\begin{center}}
\def\ec{\end{center}}
\def\bea{\begin{eqnarray}}
\def\eea{\end{eqnarray}}
\newcommand{\avg}[1]{\langle{#1}\rangle}
\newcommand{\Avg}[1]{\left\langle{#1}\right\rangle}
\title[Epidemic spreading and bond percolation  in multilayer networks]{Epidemic spreading and bond percolation  in multilayer networks}

\author{Ginestra Bianconi}

\address{School of Mathematical Sciences, Queen Mary University of London, London, E1 4NS, United Kingdom}
\ead{g.bianconi@qmul.ac.uk}
\vspace{10pt}
\begin{indented}
\item[]
\end{indented}

\begin{abstract}
The Susceptible-Infected-Recovered (SIR) model is  studied in multilayer networks with arbitrary number of links across the layers. By following the mapping to bond percolation we give the analytical expression for the epidemic threshold  and the fraction of the infected individuals in arbitrary number of layers. These results  provide   an exact prediction  of the epidemic threshold for infinite locally tree-like multilayer networks, and  an lower bound of the epidemic threshold for more general multilayer networks. The case of a multilayer network formed by two interconnected networks is specifically studied as a function of the degree distribution within and across the layers. We show that the epidemic threshold strongly depends on the degree correlations of  the multilayer structure. Finally we relate our results to the results obtained in the annealed approximation for the Susceptible-Infected-Susceptible (SIS) model.
\end{abstract}

%
%
%
%
%

\section{Introduction}

Epidemic spreading \cite{Epidemics_review,Dynamics,Doro_crit,Newman_book,Laszlo} is among the most studied processes in networks having applications spanning different disciplines from healthcare, to social sciences and finance.
Recently epidemic spreading and diffusion on multilayer networks \cite{PhysReports,Kivela,Arenas_review,diffusion,Radicchi_diff} are attracting increasing interest. In fact multilayer networks are ubiquitous and play a very important role in most spreading processes. For instance multilayer networks include transportation networks \cite{Boccaletti,Arenas_PNAS} along which viruses spread, social online networks responsible for the spread of rumors and behaviour \cite{Arenas_awareness}, and financial networks and infrastructures where cascading failures can occur \cite{Havlin1,Brummitt}.

Multilayer networks \cite{PhysReports,Kivela}  are formed by several interacting networks (called also  layers). They  can be classified in two main classes: multiplex networks and general multilayer networks.
A multilayer network is a multiplex when all the  nodes of any layer are mapped one-to-one to  all the nodes of any other layer. Additionally the links across layers only exist between corresponding nodes.
A general multilayer network instead is formed by different networks including intralinks (links within each layer) and interlink (links among different layers),  with no restrictions on where the interlinks can be placed.

So far several works have studied  the Susceptible-Infected-Suceptible (SIS) 
\cite{Arenas_awareness,Boguna_epidemics,Cozzo,Yamir,Valdano} and the Susceptible-Infected-Recovered (SIR) dynamics \cite{Stanley_SIR,Braunstein1,Immunization,Braunstein2} on multilayer networks.
The SIS model has  been studied in multiplex and multilayer networks showing that the SIS epidemics can become endemic in a multilayer network also if the single layers that form the structure could not possibly sustain the epidemics in isolation \cite{Boguna_epidemics,Cozzo}. Interestingly in \cite{Boguna_epidemics} an analytic expression of the SIS epidemic threshold has been derived in the framework of the annealed network approximation.

The effect that social online networks can have in the  actual spreading of a viral epidemic can be actually significant. This effect can be fully accounted \cite{Arenas_awareness} by a multilayer network perspective by coupling two set of SIS-like epidemics (awareness-unawareness-awareness spreading in the social online network and susceptible-infected-susceptible spreading in the social contact network). 
It is notable that  the multilayer network framework has been also successfully used   to characterize  the SIS dynamics and predict the epidemic threshold on temporal networks \cite{Valdano}.

The SIR model has been investigated extensively    on multiplex networks \cite{Braunstein1,Immunization,Braunstein2,Yagan,Yagan2,Yagan3} using the mapping to bond percolation \cite{Newman_2002}. The epidemic threshold has been predicted  and the effect of different immunization strategies investigated. Nevertheless, until now we do not have a comprehensive framework to study SIR dynamics in a general multilayer network where there is an arbitrary number of links across the different layers.

Here we provide theoretical predictions for the size of the epidemic outbreak and the epidemic threshold extending the mapping valid for the SIR dynamics on single and multiplex network to general multilayer networks.
We provide a solution of the bond percolation problem valid for locally tree-like multilayer networks.
Although percolation is widely investigated in multiplex networks \cite{Havlin1,Baxter,CDB_2016,RB1,RB2} and general multilayer networks \cite{BD1,BD2}, most of the attention has been focusing until now on percolation of interdependent networks while only few works characterize bond percolation \cite{dSouza,Makse,Cellai}. Here we are providing general results on bond percolation of multilayer networks  which  go beyond the existing literature on the subject finding closed analytical expressions for determining the percolation threshold. We are able to identify the key role of correlations in determining the percolation phase diagram of multilayer networks providing closed analytical expressions.
Finally we are able to use these results to predict the SIR epidemic threshold and the average size of the epidemic outbreak.

We note that while here we focus on the asymptotic properties of the SIR dynamics,  recent attention has been addressed to the finite time dynamical properties of the epidemic spreading models in the context of  single networks \cite{Newman_dynamics,DallAsta, Saad}. Notably significant progress in predicting and containing the epidemics  can be achieved with message passing techniques. Interestingly the apporach proposed in this paper for studying the asymptotic properties of the SIR dynamics on multilayer networks can be extended  in this direction in future publications.

\section{Mapping between the SIR epidemic spreading and bond percolation}
We consider the SIR epidemic spreading model \cite{Epidemics_review,Doro_crit,Newman_book} in which nodes can be in three possible states: susceptible, infected or recovered. Susceptible nodes can be infected if a neighbouring node is infected. Infected nodes  can spread the epidemic to neighbouring susceptible nodes with a probability that depends on the infection rate.  Recovered nodes are nodes that have been infected in the past and are effectively removed from the population, i.e. they cannot be infected but they cannot spread the epidemics either.
The SIR model is characterized by a non-equilibrium phase transition occurring as a function of the infection rate. Below the phase transition an  epidemics started from a single infected individual remains localized and dies out after infecting an infinitesimal fraction of nodes. Above the phase transition instead an epidemic outbreak is observed and the epidemics affects a finite fraction of all the nodes of the network.
On single networks the SIR dynamics is known to allow a theoretical  solution thanks to the mapping of the model to bond percolation \cite{Newman_2002,DallAsta}.
In multilayer network the SIR dynamics has been  explored using the mapping to bond percolation in the case of multiplex networks \cite{Braunstein1,Immunization,Braunstein2}. Here we provide a theoretical approach to characterize  the SIR dynamics on multilayer networks with arbitrary number of links across the layers.
We consider a multilayer network ${\cal M}$ with $M$ layers $\alpha=1,2,\ldots, M$.
Each layer $\alpha$ is formed by $N$ nodes indicated as $(i,\alpha)$ with $i=1,2,\ldots,N$.
The connections of the multilayer network are fully determined     \cite{diffusion} by the $NM\times NM$ supra-adjacency matrix $A_{i\alpha,j\beta}$ of  elements 
\bea
A_{i\alpha,j\beta}=\left\{\begin{array}{ll}1 &\mbox{if } (i,\alpha)\mbox{ is connected to } (j,\beta)\nonumber \\ 0 &\mbox{otherwise}\end{array}\right.
\eea
Every node $(i,\alpha)$ has a multilayer degree 
\bea
{\bf k}_{i\alpha}=\left(k_{i\alpha}^{[\alpha,1]},k_{i\alpha}^{[\alpha,2]},\ldots, k_{i\alpha}^{[\alpha,M]}\right)\eea
where  $k^{[\alpha,\beta]}_{i\alpha}$ indicates the number of nodes in layer $\beta$ that are connected to the node $(i,\alpha)$, i.e.
\bea
k_{i\alpha}^{[\alpha,\beta]}=\sum_{j=1}^N A_{i\alpha,j\beta}.
\eea

The SIR model over such multilayer network should describe an epidemics that spreads with different infection rates within and across the layers. To this end we indicate by  $\zeta^{[\alpha,\beta]}$ the infection rate from a node in layer $\alpha$ to a node in layer $\beta$ and we assume for simplicity that the infection rate across a single link does not depends on the direction in which the epidemic spread, i.e. $\zeta^{[\alpha,\beta]}=\zeta^{[\beta,\alpha]}$.
We indicate by $\mu$ the rate a which an infected node recovers. 
With this notation, we can derive, following similar steps used for the SIR dynamics on single layers \cite{Newman_2002} the value of the transmissibility $T^{[\alpha,\beta]}$ of the infection across a link going from a node in layer $\alpha$ to a node in layer $\beta$.  This is equal to the probability that an infected node in layer $\alpha$, neighbour  of a susceptible node in layer $\beta$, actually transmits the epidemic to that node. 
Since every infected individual recover at constant rate $\mu$, the distribution $P(\tau)$ of the  lifetime  $\tau$  of an  infected node is Poisson  
\bea
P(\tau)=\mu e^{-\mu \tau}.
\eea
Indicating with $T^{[\alpha,\beta]}_{\tau}$ the probability that an infected node with lifetime $\tau$ transmits the infection to a neighbouring susceptible node, we have 
\bea
T^{[\alpha,\beta]}=\int d\tau P(\tau)T_{\tau}^{[\alpha,\beta]}.
\label{T1}
\eea
Given that the rate at which an infected individual in layer $\alpha$ infects a susceptible individual in layer $\beta$ is constant and given by $\zeta^{[\alpha,\beta]}$, we have 
\bea
T_{\tau}^{[\alpha,\beta]}=1-\exp{\left[-\zeta^{[\alpha,\beta]}\tau\right]}
\eea
Finally by performing the integral in Eq. $(\ref{T1})$ we obtain the transmissibility $T^{[\alpha,\beta]}$ 
\bea 
T^{[\alpha,\beta]}=\frac{\lambda^{[\alpha,\beta]}}{1+\lambda^{[\alpha,\beta]}},
\eea
where $\lambda^{[\alpha,\beta]}=\zeta^{[\alpha,\beta]}/\mu$.

Having calculated the value of the transmissibility $T^{[\alpha,\beta]}$ within and across the layers, we can use the well known result that maps the SIR dynamics to bond percolation in a single network \cite{Newman_2002} and extend it directly to a generic multilayer network. Therefore the SIR dynamics on a generic multilayer network maps to bond percolation where the probability $p_{\alpha\beta}$ to retain a link going from a node in layer $\alpha$ to a node in layer $\beta$ is given by 
\bea
p_{\alpha\beta}=T^{[\alpha,\beta]}=\frac{\lambda^{[\alpha,\beta]}}{1+\lambda^{[\alpha,\beta]}}.
\label{perSIR}
\eea
Note that since we have assumed that the infection rate is symmetric $\zeta^{[\alpha,\beta]}=\zeta^{[\beta,\alpha]}$ we have also  $p_{\alpha\beta}=p_{\beta\alpha}$.
In this mapping the percolation cluster represent the set of recovered nodes at the end of the epidemics outbreak. Therefore the size of the giant component indicates the size of the outbreak.
Consequently the epidemic threshold corresponds to the epidemic threshold of the bond percolation problem.

In this way the study of  SIR model on multilayer networks is fully reduced to bond percolation  on the same network structure.
Bond percolation on multilayer networks is an interesting critical phenomenon in itself characterizing the robustness of the network to random damage.

Here we show that  bond percolation in multilayer networks with arbitrary number of links across the layers, can be theoretically solved as long as the multilayer network is locally tree-like. This solution will reveal the important effects of correlations in the properties of epidemic spreading in multilayer networks.

\section{SIR model and bond percolation in a single multilayer network}

\subsection{Bond percolation in a generic multilayer network}

Given a  locally tree-like multilayer network of $M$ layers where links between nodes of layer $\alpha$ and layer $\beta$ are retained with probability $p_{\alpha\beta}$, it is possible to derive the expression of the average size of the giant component using a message passing algorithm.
This approach is a straightforward  generalization of the message passing algorithm use to detect the giant component in single networks \cite{Lenka,Mezard,DallAsta,Hamilton}.
In fact, the only difference is that the probability to retain a link is now dependent on the classification of nodes in different layers.
In this algorithm the probability $\sigma_{i\alpha}$ that node $(i,\alpha)$ is in the giant component depends on a set of messages exchanged between neighbouring nodes whose values are determined by a 
self-consistent set of equations.

Let us  indicate with $\sigma_{i\alpha \to j\beta}$ the generic message sent from a node $(i,\alpha)$ to a neighbouring node $(j,\beta)$ and representing the probability that node $(i,\alpha)$ connects node $(j,\beta)$ to the giant component whereas the link between node $(i,\alpha)$ and node $(j,\beta)$ has not been initially damaged.
Therefore $\sigma_{i\alpha,\to j\beta}$ indicates the probability that node $(i,\alpha)$ is connected by a non-damaged link to at least one  node $(\ell,\gamma) \neq (j,\beta)$ which connects it to other nodes beloging to the giant component.
This recursive algorithm defines the  message passing equations satisfied by the messages $\sigma_{i\alpha\to j\beta}$ on   locally tree-like multilayer networks: 
\bea
\sigma_{i\alpha\to j\beta}=1-\prod_{(\ell,\gamma)\in  N(i,\alpha)\setminus (j,\beta)}(1-p_{\gamma \alpha}\sigma_{\ell \gamma \to i\alpha}),
\label{per}
\eea
where $N(i,\alpha)$ indicates the set of nodes that are neighbour of node $(i,\alpha)$ within the same layer or across different layers.
The probability $\sigma_{i\alpha}$ that a node $(i,\alpha)$ belongs to the giant component is given by the probability that node $(i,\alpha)$ has at least a  non-damaged connection to   a node $(\ell,\gamma)$ that connects it to the giant component, i.e.
\bea
\sigma_{i\alpha}=1-\prod_{(\ell,\gamma)\in  N(i,\alpha)}(1-p_{\gamma\alpha}\sigma_{\ell \gamma \to i\alpha})
\label{per2}
\eea
Finally the average fraction of nodes in   the giant component is given by 
\bea
S=\frac{1}{MN}\sum_{i=1}^N \sum_{\alpha=1}^M\sigma_{i\alpha}.
\eea

By using the   mapping between the SIR dynamics and percolation, $S
$ can  also be interpreted as the size of the epidemic  outbreak when $p_{\alpha \beta}$ is related to the infection rates $\lambda^{[\alpha,\beta]}$ according to Eq. $(\ref{perSIR})$.

The message passing Eqs. $(\ref{per})$ and $(\ref{per2})$ have always a trivial solution $\sigma_{i\alpha}=\sigma_{i\alpha \to j\beta}=0$.
However for large enough percolation probabilities $p_{\alpha\beta}$ they develop a non-trivial solution consistent with a non-vanishing  fraction of nodes in the giant component $S$.

The percolation threshold indicate the values of $p_{\alpha\beta}$ where  this transition occurs. 
It  can be found by linearizing the message passing equations close to the trivial solution $\sigma_{i\alpha \to j \beta}=0$. Writing $\sigma_{i\alpha \to j \beta}=\epsilon_{i\alpha \to j \beta}\ll 1$ the linearized Eq. $(\ref{per})$ reads
\bea
\epsilon_{i\alpha \to j\beta}=\sum_{(\ell,\gamma)\in N(i,\alpha)\setminus(j,\beta)}
p_{\gamma \alpha}\epsilon_{\ell \gamma \to i\alpha}.
\eea
This equation can be written in matrix form as 
\bea
{\bf \epsilon}={\bf B}{\epsilon}
\label{eMe}
\eea
where ${\bf B}$ is the non-backtracking matrix of the multilayer network. It is  a $L\times L$ matrix where $L$ indicates the total number of links in the multilayer network and has elements 
\bea
B_{i\alpha \to j\beta;\ell \gamma \to m\delta}=p_{\gamma\alpha}\left[1-\delta_{(\ell,\gamma),(j,\beta)}\right]\delta_{(m,\delta),(i,\alpha)}.
\eea
From Eq. $(\ref{eMe})$ it follows immediately that a small perturbation of the messages from the trivial solution $\sigma_{i\alpha\to j\beta}=0$ is suppressed if the maximum eigenvalue $\Lambda_B$ of the non-backtracking matrix  is smaller than one.
Nevertheless of $\Lambda_B>1$ the perturbation is enhanced corresponding to the onset of the instability for the trivial solution $\sigma_{i\alpha\to j\beta}=0$.
It follows that we will observe an epidemic in a  given finite network for values of the probabilities $ p_{\alpha\beta}$ such that the maximum eigenvalue $\Lambda_B$ of the non-backtracking matrix is greater than one, i.e.
\bea
\Lambda_{B}>1.
\label{lambdaM}
\eea
By relating $p_{\alpha\beta}$  to the infection rates of the SIR model according to Eq. $(\ref{perSIR})$, it is straightforward to use  Eq. $(\ref{lambdaM})$ to determine  the epidemic threshold of the SIR model by imposing $\Lambda_B=1$.
Note that the maximum eigenvalue of the non-backtracking matrix is guaranteed  to provide the exact percolation and hence epidemic threshold only for locally tree-like networks of infinite sizes. For finite networks with loops it provides a lower-bound to the epidemic threshold.

\section{SIR model and bond percolation  in an esemble of  multilayer networks}

\subsection{Multilayer networks with any number of layers $M$}

While the message passing approach is very general as it applies to any given multilayer network, its results do not explicitly determine the effect of the multilayer structure in determining the critical properties of the bond percolation transition and consequently of the epidemic spreading transition.

Analytical insights on the important role of multiplexity can  instead be gained by studying these processes on multilayer network ensembles including a controlled level of multilayer degree  correlations. 

Additionally, it is well knwon in statistical physics that phase transitions are well defined only in the infinite network limit. 
Therefore, characterizing the epidemic spreading in multilayer network ensembles, allows to explore  the properties of this non-equilibrium phase transition.

For simplicity, we consider here  a multilayer ensemble with given multilayer degree sequence (an extension to more general multilayer ensembles is given in the appendix). In this case   the  probability of a generic multilayer network ${\cal M}$ is given by 
\bea
P({\cal M})=\frac{1}{Z}\delta\left(k_{i\alpha}^{[\alpha,\beta]},\sum_{j=1}^N A_{i\alpha,j\beta}\right),
\eea
where $\delta(x)$ is the Kronecker delta, and $Z$ is a normalization factor.
The multilayer degree sequence is choosen in such a way that the probability ${\pi}_{i\alpha,j\beta}$ of a link between node $(i,\alpha)$ and node $(j,\beta)$ follows the simple product rule 
\bea
{\pi}_{i\alpha,j\beta}=\frac{k_{i\alpha}^{[\alpha,\beta]}k_{j\beta}^{[\beta,\alpha]}}{\Avg{k^{[\alpha,\beta]}}N}, 
\eea
i.e. we assume that there are no degree-degree correlations.
This expression in ensured by  the degree cutoffs $K^{[\alpha,\beta]}=\max_{i} k_{i,\alpha}^{[\alpha,\beta]}$ satisfying the following set of relations
\bea
\frac{K^{[\alpha,\beta]}K^{[\beta,\alpha]}}{\Avg{k^{[\alpha,\beta]}}N}<1.
\eea
Additionally we note here that the multilayer degrees must necessarily satisfy 
\bea
\Avg{k^{[\alpha,\beta]}}=\Avg{k^{[\beta,\alpha]}}.
\eea
In fact, since each layer has the same number of nodes, this implies that  the number of links going from layer $\alpha$ to layer $\beta$ is equal to the  number of links going from layer $\beta$ to layer $\alpha$.
We indicate with $P_{\alpha}({\bf k})$ the probability that a generic node $(i,\alpha)$ of layer $\alpha$ has multilayer degree ${\bf k}_{i\alpha}={\bf k}$, i.e.
\bea
P_{\alpha}({\bf k})=\frac{1}{N}\sum_{i=1}^N \delta\left({\bf k}_{i\alpha},{\bf k}\right).
\eea
The equations determining the  fraction of nodes in the giant component of this ensemble of multilayer network can be obtained by averaging the messages going from  one layer to another layer.
The average messages $S_{\alpha,\beta}^{\prime}=\Avg{\sigma_{i\alpha\to j\beta}}$ indicate the probability that by following a link we reach a node that is in the giant component.
Therefore the average messages $S_{\alpha,\beta}^{\prime}$ 
satisfy the following set of equations
\bea
S_{\alpha\beta}^{\prime}&=&x_{\alpha \beta}\left[1-\sum_{{\bf k}}\frac{k^{[\alpha,\beta]}}{\Avg{k^{[\alpha,\beta]}}}P_{\alpha}({\bf k})\prod_{\gamma}(1-p_{\gamma\alpha}S_{\gamma \alpha}^{\prime})^{k^{[\alpha,\gamma]}-\delta(\gamma,\beta)}\right],
\label{Spab}
\eea
where  $x_{\alpha\beta}=1$ if there is at least one connection between layer $\alpha$ and layer $\beta$, otherwise $x_{\alpha,\beta}=0$, i.e.
\bea
x_{\alpha\beta}=1-\delta\left(0,\Avg{k^{[\alpha,\beta]}}\right).
\eea
In Eq. ($\ref{Spab}$) we have adopted the notation $S_{\alpha\beta}^{\prime}=0$ whereas $x_{\alpha\beta}=0$.
The probability $S_{\alpha}=\Avg{\sigma_{i\alpha}}$ that a generic node $(i,\alpha)$ of layer $\alpha$ is in the giant component is expressed in terms of the average messages $S_{\alpha\beta}^{\prime}$ as
\bea
S_{\alpha}=1-\sum_{{\bf k}}P_{\alpha}({\bf k})\prod_{\gamma}(1-p_{\gamma\alpha}S_{\gamma \alpha}^{\prime})^{k^{[\alpha,\gamma]}}.
\label{S}
\eea
Finally the fraction of nodes in the giant component is given by 
\bea
S=\frac{1}{M}\sum_{\alpha=1}^M S_{\alpha}.
\eea

The Eqs. $(\ref{Spab}),(\ref{S})$ generalize the well known equations determining the size of the giant componet   on single networks in the locally tree-like approximation (see for instance Ref. \cite{Doro_crit}).

The  percolation threshold is found by linearizing this equation close to the trivial solution $S^{\prime}_{\alpha,\beta}=0$ obtaining the system of equations
\bea
S^{\prime}_{\alpha\beta}=\sum_{\gamma}p_{\gamma\alpha}x_{\alpha\beta}\frac{\Avg{k^{[\alpha,\beta]}[k^{[\alpha,\gamma]}-\delta(\gamma,\beta)]}}{\Avg{k^{[\alpha,\beta]}}}S^{\prime}_{\gamma\alpha}.
\label{L1}
\eea
which can be written as 
\bea
{\bf S}^{\prime}={\bf J}{\bf S}^{\prime}
\eea
where ${\bf J}$ is the $M^2\times M^2$ Jacobian matrix of the system of Eqs.  $(\ref{L1})$ of elements
\bea
J_{\alpha\beta;\gamma\alpha}=p_{\gamma\alpha}x_{\alpha\beta}\frac{\Avg{k^{[\alpha,\beta]}[k^{[\alpha,\gamma]}-\delta(\gamma,\beta)]}}{\Avg{k^{[\alpha,\beta]}}}.
\label{J1a}
\eea
Note that in Eq. $(\ref{L1})-(\ref{J1a})$ the have  adopted the following notation: whereas $\Avg{k^{[\alpha,\beta]}}=0$ we  take $x_{\alpha\beta}\frac{\Avg{k^{[\alpha,\beta]}[k^{[\alpha,\gamma]}-\delta(\gamma,\beta)]}}{\Avg{k^{[\alpha,\beta]}}}=0$.
Above the  transition, this system must develop a set of non trivial solutions.
Therefore the transition point  is obtained by imposing that  
 the maximum eigenvalue $\Lambda_J$ of the matrix ${\bf J}$ satisfies
\bea
\Lambda_J=1.
\eea
\subsection{Multilayer  networks with $M=2$ layers}
Let us consider the case of a duplex network.
The system of linearized Eqs. $(\ref{L1})$ reads 
\bea
S^{\prime}_{11}&=&\kappa_{11} p_{11}S^{\prime}_{11}+{\cal K}_{12}p_{12} S^{\prime}_{21}\nonumber \\
S^{\prime}_{22}&=&\kappa_{22}p_{22} S^{\prime}_{22}+{\cal K}_{21}p_{12} S^{\prime}_{12}\nonumber \\
S^{\prime}_{12}&=&{\cal W}_{12}p_{11} S^{\prime}_{11}+{\kappa}_{12} p_{12}S^{\prime}_{21}\nonumber \\
S^{\prime}_{21}&=&{\cal W}_{21} p_{22}S^{\prime}_{22}+{\kappa}_{21}p_{12} S^{\prime}_{12}\nonumber \\
\eea
 where 
 \bea
 \kappa_{\alpha\beta}&=&x_{\alpha\beta}\frac{\Avg{k^{[\alpha, \beta]}(k^{[\alpha, \beta]}-1)}}{\Avg{k^{[\alpha,\beta]}}}\nonumber \\
 {\cal K}_{12}&=&x_{11}\frac{\Avg{k^{[1,1]}k^{[1,2]}}}{\Avg{k^{[1,1]}}}\nonumber \\
  {\cal K}_{21}&=&x_{22}\frac{\Avg{k^{[2,2]}k^{[2,1]}}}{\Avg{k^{[2,2]}}}\nonumber \\
  {\cal W}_{12}&=&x_{12}\frac{\Avg{k^{[1,2]}k^{[1,1]}}}{\Avg{k^{[1,2]}}}\nonumber \\
  {\cal W}_{21}&=&x_{12}\frac{\Avg{k^{[2,1]}k^{[2,2]}}}{\Avg{k^{[2,1]}}} 
 \eea
 The  transition is therefore obtained when the following condition is satisfied:
  \bea
 0&=& (1-p_{11}\kappa_{11})(1-p_{22}\kappa_{22})-p_{12}^2{\cal R}_{12},
 \label{pm}
 \eea
 where 
 \bea
{\cal R}_{12}&=& \left(p_{11}{\cal W}_{12}{\cal K}_{12}+\kappa_{12}-p_{11}\kappa_{12}\kappa_{11}\right)\nonumber\\ 
&&\times
  \left(p_{22}{\cal W}_{21}{\cal K}_{21}+\kappa_{21}-p_{22}\kappa_{21}\kappa_{22}\right)
  \label{R12}
  \eea
In the following we discuss few specific limiting behaviour that can be considered starting from these equations.
\begin{itemize}
\item
{\em Only interlayer connectivity-}
  For the case in which only the links within each layer exist, i.e. $x_{12}=0$ or $p_{12}=0$, we  recover the  percolation transitions of the single layers \cite{Newman_book,Doro_crit} 
  \bea
  p_{\alpha\alpha}\kappa_{\alpha\alpha}=1,
  \eea
  or equivalently,
  \bea
  p_{\alpha \alpha}=\frac{\Avg{k^{[\alpha,\alpha]}}}{\Avg{k^{[\alpha,\alpha]}(k^{[\alpha,\alpha]}-1)}}.
  \eea
  Therefore using Eq. $(\ref{perSIR})$ we obtain the epidemic threshold \cite{Newman_book,Doro_crit,Newman_2002} 
  \bea
  \lambda^{[\alpha,\alpha]}=\frac{\Avg{k^{[\alpha,\alpha]}}}{\Avg{(k^{[\alpha,\alpha]})^2}-2\Avg{k^{[\alpha,\alpha]}}}.
  \eea
  \item
  {\em Only intralayer connectivity-}
  For the case that only the interlinks exist $x_{11}=x_{22}=0$ or $p_{11}=p_{22}=0$ we obtain for the bond percolation transition of bipartite networks \cite{Newman_2002}, 
  \bea
 p_{12}^2 \kappa_{12}\kappa_{21}=1,
  \eea
  or equivalently
  \bea
  p_{12}&=&\sqrt{\frac{1}{\kappa_{12}\kappa_{21}}}\nonumber \\
  &=&\sqrt{\frac{\Avg{k^{[1,2]}}}{\Avg{k^{[1,2]}(k^{[1,2]}-1)}}\frac{\Avg{k^{[2,1]}}}{\Avg{k^{[2,1]}(k^{[2,1]}-1)}}}
  \eea
  which can be vanishingly small for an heterogeneous distribution of either of the two   $k^{[1,2]}$ or $k^{[2,1]}$.  
  Interestingly in this case the multilayer network might have a giant component also if one layer (for example layer 2) has $\kappa_{21}<1$ provided that $\kappa_{12}\kappa_{21}>1$.
  For example one can have one layer (layer 1) with scale-free distribution of the intralayer degrees $k^{[1,2]}$ and one layer (layer 2) with a Poisson degree distribution $k^{[2,1]}$ with average degree smaller than 1 and still having a giant component. 
  Given the found epidemic threshold it can be easily deduced using Eq. $(\ref{perSIR})$ that the  SIR epidemic threshold is given by  \cite{Newman_2002} 
  \bea
  \lambda^{[1,2]}&=&\frac{1}{\sqrt{\kappa_{12}\kappa_{21}}-1}. \\
  \eea
  
  \item{\em Both interlayer and intralayer connectivity- }
  In general we have that the relation between $p_{12}$ at the transition point depends on the values of  $p_{11}$ and $p_{22}$.
  As long as 
  \bea
  p_{\alpha\alpha}\leq\frac{1}{\kappa_{\alpha\alpha}},
  \eea
  for $\alpha=1,2$ the percolation threshold satisfies
    \bea
  p_{12}=\sqrt{\frac{(1-p_{11}\kappa_{11})(1-p_{22}\kappa_{22})}{{\cal R}_{12}}}.
  \label{pt}
  \eea 
  Using Eq. $(\ref{perSIR})$ we obtain that the SIR epidemic threshold satisfies
  \bea
  \lambda^{[1,2]}=\frac{\sqrt{{\cal Q }}}{\sqrt{{\cal R}_{12}}-\sqrt{\cal Q }},
  \label{lc}
  \eea
  where
  \bea
{ \cal Q}=\left(1-\frac{\lambda^{[1,1]}}{1+\lambda^{[1,1]}}\kappa_{11}\right)\left(1-\frac{\lambda^{[2,2]}}{1+\lambda^{[2,2]}}\kappa_{22}\right) 
  \eea 
  and
  \bea
  \frac{\lambda^{[\alpha,\alpha]}}{1+\lambda^{[\alpha,\alpha]}}\kappa_{\alpha\alpha}\leq 1.
  \eea
  \end{itemize}
  The term ${\cal R}_{12}$ given by Eq. $(\ref{R12})$ is dependent on ${\cal W}_{12},{\cal W}_{21}$ and ${\cal K}_{12}, {\cal K}_{21}$ that evaluate the  (normalized) correlations between the interlayer degree and the intralayer degree. As a consequence of this, both the epidemic threshold given by Eq. $(\ref{pt})$ and the epidemic threshold given by Eq.$(\ref{lc})$ are not only strongly dependent  on the presence of broad interdegree and intradegree distributions but they are also significantly affected by the correlations between the interdegrees and the intradegrees.

  \subsection{Effect of interdegree and intradegree correlations}
  
\begin{figure}[htb]
\begin{center}
	\includegraphics[width=0.97\columnwidth]{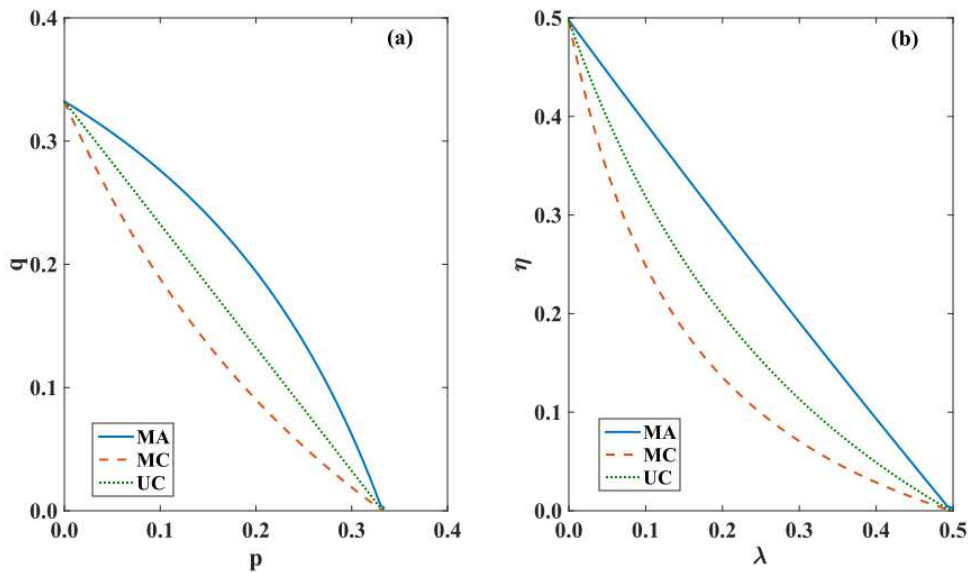}	
	\caption{(Color online) Percolation threshold (panel a) and epidemic threshold (panel b) for maximally correlated (MC), maximally anticorrelated (MA) or uncorrelated (UC) interdegree and intradegree sequences. The  multilayer networks have  identical interlayer degree sequences and intralayer degree sequences  drawn from a Poisson distribution with average degree $c=3$. The network size is  $N=10^4$ nodes.  }
	\label{fig:corr}
\end{center}
\end{figure}

  Consider a multilayer network with $M=2$ layers.
 Let us assume to have  a given multidegree distribution of interlayer degrees and intralayer degrees, and let us consider the effect of  changing the correlations between interlayer and intralayer degrees.\\
 For high positive correlations of interlayer and intralayer degrees ${\cal R}_{12}$ is higher and the epidemic threshold of the network is smaller. Additionally in this limit the multilayer network is more robust and the percolation threshold is smaller.\\
 For large anticorrelations of interlayer and intralayer degrees ${\cal R}_{12}$ is smaller and the epidemic threshold of the network is larger. Additionally in this limit the multilayer network is more fragile and the percolation threshold is larger.\\\
 To show the effect of degree correlations in determining the percolation threshold and the epidemic threshold, we considered a network with identical interlayer degree sequences and intralayer degree sequences, in which the intralayer degree and the interlayer degree of each node are either maximally positively correlated (MC), maximally anti-correlated (MA) or uncorrelated (UC).
 These correlated multilayer networks are constructed starting from multilayer network with the same interlayer network structure by changing the way the intralinks are placed. For maximally positive correlated (MC) multilayer networks  the intradegree and the interdegree sequences are first sorted in descending order.  To each node of rank $r$ in the intradegree sequence it is   assigned the interdegree having  the same rank $r$. Subsequently the bipartite network between the two layers is randomly drawn by preserving the interdegree of each node. For maximally anticorrelated (MA) networks  we proceed as in the previous case with the exception that the intradegree and interdegree distribution are sorted in opposite order (one sequence in increasing order and the other sequence in decresing order). Finally for the case of uncorrelated (UC) multilayer networks the interdegree is assigned randomly to any node of a given layer of the multilayer network by performing a random permutation of the corresponding interdegree sequence.
In panel (a) of figure $\ref{fig:corr}$ we show the percolation threshold $q=q(p)$ for  Poisson multilayer networks where we have  indicated with $p=p_{11}=p_{22}$ and $q=p_{12}=p_{21}$ the probability to retain respectively the interlinks and the intralinks.
In panel (b) of figure $\ref{fig:corr}$ we show the epidemic threshold $\eta=\eta(\lambda)$ for the same networks where we have  indicated with $\lambda=\lambda^{[1,1]}=\lambda^{[2,2]}$ and $\eta=\lambda^{[1,2]}=\lambda^{[2,1]}$ the infectivity  of  interlinks and the intralinks.
\subsection{Comparison of the epidemic threshold of the SIR and  SIS models}
It might be instructive to compare the epidemic threshold of the SIR model with the epidemic threshold of the SIS model as predicted by the annealed network approximation.
We specifically compare the results obtained here for the SIR dynamics, exact in the limit of a locally tree-like network of inifite size, with the results obtained for the SIS dynamics using the annealed approximation in Ref. \cite{Boguna_epidemics} that constitute an approximation even for infinite locally tree-like networks.  
Summing up the results of the previous section, we have found that in a multilayer network with two layers the SIR epidemic threshold is occurring
when the following condition is satisfied:
  \bea
 0&=& \left(1-\frac{\lambda^{[1,1]}}{1+\lambda^{[1,1]}}\kappa_{11}\right)\left(1-\frac{\lambda^{[2,2]}}{1+\lambda^{[2,2]}}\kappa_{22}\right)- \left(\frac{\lambda^{[1,2]}}{1+\lambda^{[1,2]}}\right)^2{\cal R}_{12},
 \eea
 where 
 \bea
{\cal R}_{12}&=& \left(\frac{\lambda^{[1,1]}}{1+\lambda^{[1,1]}}{\cal W}_{12}{\cal K}_{12}+\kappa_{12}-\frac{\lambda^{[1,1]}}{1+\lambda^{[1,1]}}\kappa_{12}\kappa_{11}\right)\nonumber\\ 
&& \times
  \left(\frac{\lambda^{[2,2]}}{1+\lambda^{[2,2]}}{\cal W}_{21}{\cal K}_{21}+\kappa_{21}-\frac{\lambda^{[2,2]}}{1+\lambda^{[2,2]}}\kappa_{21}\kappa_{22}\right).
  \label{R12b}
  \eea
In fact these equations can be directly obtained by Eqs. $(\ref{pm})$ and $(\ref{R12})$ using the expression of $p_{\alpha,\beta}$ in terms of the infectivities $\lambda^{[\alpha,\beta]}$ given by Eq. $(\ref{perSIR})$.

On the other side,  the annealed approximation of the SIS model on the same multilayer network predicts \cite{Boguna_epidemics}   the epidemic spreading transition for 
    \bea
 0&=& \left[1-\lambda^{[1,1]}(\kappa_{11}+1)\right]\left[1-\lambda^{[2,2]}(\kappa_{22}+1)\right]- \left(\lambda^{[1,2]}\right)^2\hat{\cal R}_{12},
 \eea
 where 
 \bea
\hat{\cal R}_{12}&=& \left[\lambda^{[1,1]}{\cal W}_{12}{\cal K}_{12}+\kappa_{12}+1-\lambda^{[1,1]}(\kappa_{12}+1)(\kappa_{11}+1)\right]\nonumber\\ 
&&
 \times\left[\lambda^{[2,2]}{\cal W}_{21}{\cal K}_{21}+\kappa_{21}+1-\lambda^{[2,2]}(\kappa_{21}+1)(\kappa_{22}+1)\right].
  \eea
  
  Therefore the  SIR equations for the epidemic threshold  reduce to the  equations for the epidemic threshold  of the  SIS model obtained with the annealed approximation  up to the set of substitutions 
 \begin{equation}
 \begin{array}{rcl}
 \kappa_{\alpha\beta} &\to  &\kappa_{\alpha\beta}+1,\nonumber \\
\frac{\lambda^{[\alpha,\beta]}}{1+\lambda^{[\alpha,\beta]}} &\to &\lambda^{[\alpha,\beta]}.
\end{array}
 \end{equation}
These results  generalize similar results obtained for the case of single networks \cite{Newman_book,Doro_crit,Dynamics,Epidemics_review}

\section{Numerical results in single multilayer networks}
To check the proposed theory with the simulation results of the SIR epidemic spreading, we have considered  multilayer networks with two layers ($M=2$).
We provide evidence than in general multilayer networks we observe the same qualitative phenomenon observed for the SIS dynamics: mainly that the SIR epidemics can spread also if the single layers cannot sustain the epidemics when taken in isolation.

\begin{figure}[htb]
\begin{center}
	\includegraphics[width=0.97\columnwidth]{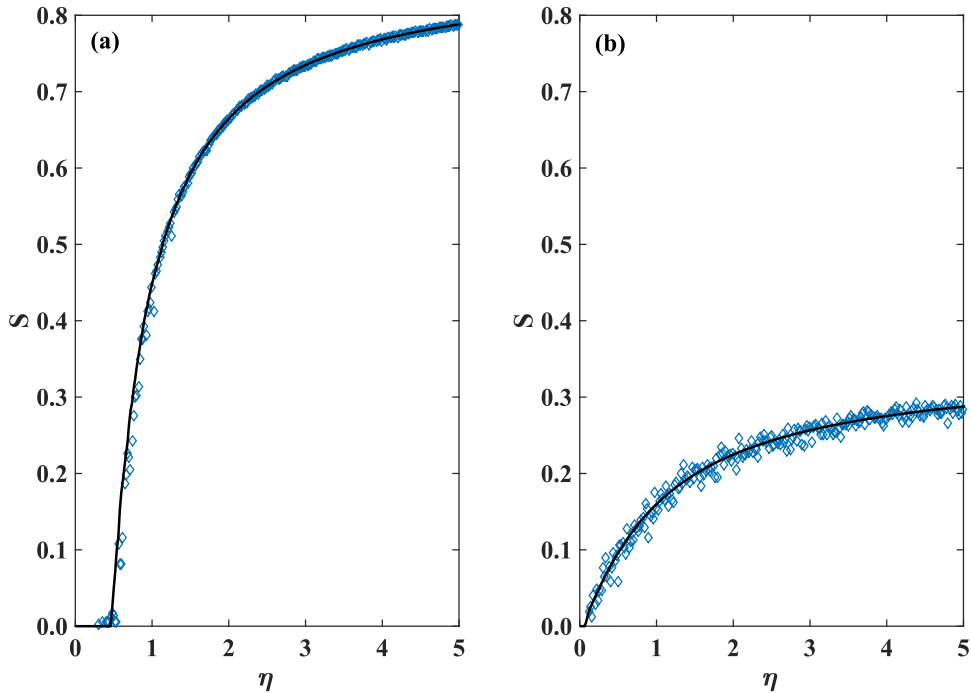}	
	\caption{(Color online) Size of the averarge epidemic outbreak $S$ as function of $\eta=\lambda^{[1,2]}=\lambda^{[2,1]}$. In panel (a) we show the case of a multilayer with Poisson intradegree and interdegree distribution with respectivelly average degree one and two. In this case $\lambda=\lambda^{[1,1]}=\lambda^{[2,2]}=0.5$. In panel (b) we show the case of a multilayer network including only interlinks. The interdegree distribution is scale-free with exponent $\gamma=2.1$ for one layer and Poisson for the other layer. The  average interdegree is below one $\Avg{k^{[1,2]}}=\Avg{k^{[2,1]}}=0.9$.
The network size is  $N=10^4$ nodes for both panel (a) and panel (b). The simulation results are averaged over 100 runs.  }
	\label{fig:sim}
\end{center}
\end{figure}

To this end, we have considered two different  multilayer networks. 
In the first case, (see panel (a) figure $\ref{fig:sim}$) the intralayer degree distribution is Poisson with average degree one while  the interlayer degree distribution is Poisson with average degree two. The intralayer infectivity $\lambda=\lambda^{[1,1]}=\lambda^{[2,2]}$ is set at a constant value $\lambda=0.5$ and the average size of the  epidemic outbreak is measured as a function of the interlayer infectivity $\eta=\lambda^{[1,2]}=\lambda^{[2,1]}$.
Since $p_{11}=p_{22}=1/3$ the single layers formed by Poisson networks with average degree one , i.e. $\kappa_{11}=\kappa_{22}=1$, cannot sustain the epidemics. Nevertheless as $\eta$ increases we observe  epidemic outbreaks.
 
In the second case (see panel (b) figure $\ref{fig:sim}$) the multilayer network includes only interlinks across the two layers. One layer has a very skewed  scale-free interdegree distribution (power-law distribution with exponent $\gamma=2.1$), the other layer has a Poisson interdegree distribution with average  smaller than one.
Here also we observe that as a function of $\eta=\lambda^{[1,2]}=\lambda^{[2,1]}$ it is possible to observe an epidemic spreading.
We notice that this occur even if even if the average interdegree of one layer is Poisson with average degree smaller than one, because the intradegree distribution of the other layer is sufficiently broad.

In both cases the  simulations of the SIR epidemic spreading   match very well the predictions obtained by  solving the corresponding bond percolation problem using the message passing technique.
Note $S$ indicates the average size of the epidemic outbreak, therefore events that do not span a finite fraction of the network are disregarded, as these  might  correspond to epidemics starting from small connected components of the multilayer network. 
 
  \section{Conclusions}
  
  In conclusion in this paper we have studied the SIR epidemics in a general multilayer network with arbitrary number of links across the layers.
  Using a well established framework to study the SIR dynamics \cite{Newman_2002}, we have mapped the problem to bond percolation in multilayer networks.
 We have shown that the message passing technique applied to the bond percolation problem  is a very powerful method to predict the average size of epidemic outbreaks as long as the multilayer network is locally tree-like.
 The characterization of the phase diagram of the SIR dynamics in an ensemble of multilayer networks has shown how the degree correlations between interlinks and intralinks can anticipate or posticipate  the onset of an epidemic outbreak.
 Finally we have related the epidemic threshold of the SIR model obtained here  thanks to the mapping to bond percolation with the annealed approximation predictions of the epidemic spreading of the SIS model on multilayer networks. 
 We hope that our work could open the venue for further investigations of the SIR dynamics in multilayer networks exploring the wide range of disciplines where
 this model is relevant including biology,  social science and finance.
  
\section{Acknowledgments}
The Author acknowledges interesting discussions with Filippo Radicchi.

  \appendix
   \section{Generalization to degree-degree correlated multilayer networks}
The multilayer network ensemble that we consider here includes all the multilayer networks ${\cal M}$ with given sequence of multilayer degree $\{{\bf k}_{i\alpha}\}_{i=1,2,\ldots, N;\alpha=1,2,\ldots, M}$ 
with degree-degree correlations.
Let us  indicate with $\tilde{P}_{\beta,\alpha}({\bf k}|\hat{\bf k})$ the probability that a link of a node in layer $\beta$ with multilayer degree $\hat{\bf k}$ connects that node to a node in layer $\alpha$ with multilayer degree ${\bf k}$. Additionally let us indicate with $P_{\alpha}({\bf k})$ the probability that a node in layer $\alpha$ has degree ${\bf k}$.
With this notation we can write the closed set of equations determining  the average message  $\tilde{S}_{\alpha\beta}^{\prime}(\hat{\bf k})$ sent from a node in layer $\alpha$ to a node in layer $\beta$ with degree $\hat{\bf k}$, if the link between the node $(i,\alpha)$ and the node $(j,\alpha)$ is not removed i.e. $\tilde{S}_{\alpha\beta}^{\prime}(\hat{\bf k})=\Avg{\sigma_{i\alpha\to j\beta}|{\bf k}_{j\beta}=\hat{\bf k}}$
which reads 
\bea
\tilde{S}_{\alpha\beta}^{\prime}(\hat{\bf k})&=&x_{\alpha \beta}\left[1-\sum_{{\bf k}}\tilde{P}_{\beta\alpha}({\bf k}|\hat{\bf k})\right.\nonumber \\
&&\hspace{-5mm}\left.\times\prod_{\gamma}(1-p_{\gamma\alpha}\tilde{S}_{\gamma \alpha}^{\prime}({\bf k}))^{{k}^{[\alpha,\gamma]}-\delta(\gamma,\beta)}\right].\nonumber
\label{SpC}
\eea
Here  $x_{\alpha\beta}=1$ if there is at least one connection between layer $\alpha$ and layer $\beta$, otherwise $x_{\alpha,\beta}=0$, i.e.
\bea
x_{\alpha\beta}=1-\delta\left(0,\Avg{k^{[\alpha,\beta]}}\right).
\eea

The probability $\tilde{S}_{\alpha}=\Avg{\sigma_{i\alpha}}$ that a generic node $(i,\alpha)$ of layer $\alpha$ is in the giant component is expressed in terms of the average messages $\tilde{S}_{\alpha\beta}^{\prime}({\bf k})$ as
\bea
\tilde{S}_{\alpha}=1-\sum_{{\bf k}}P_{\alpha}({\bf k})\prod_{\gamma}(1-p_{\gamma\alpha}\tilde{S}_{\gamma \alpha}^{\prime}({\bf k}))^{k^{[\alpha,\gamma]}}.
\label{SC}
\eea
Finally the fraction of nodes in the giant component is given by 
\bea
S=\frac{1}{M}\sum_{\alpha=1}^M \tilde{S}_{\alpha}.
\eea
The  percolation threshold is found by linearizing this equation close to the trivial solution $\tilde{S}^{\prime}_{\alpha,\beta}({\bf k})=0$ obtaining the system of equations
\bea
\tilde{S}^{\prime}_{\alpha,\beta}(\hat{\bf k})=\sum_{\gamma}p_{\gamma\alpha}x_{\alpha\beta}\sum_{\hat{\bf k}}\tilde{P}_{\beta\alpha}({\bf k}|\hat{\bf k})\left[{k}^{[\alpha,\gamma]}-\delta(\gamma,\beta)\right]\tilde{S}^{\prime}_{\gamma\alpha}({\bf k}).\nonumber\\
\label{SpC}
\eea
which can be written as 
\bea
{\bf \tilde{S}}^{\prime}=\hat{\bf J}{\bf \tilde{S}}^{\prime}
\eea
where ${\bf J}$ is the $M^2P\times M^2P$ Jacobian matrix (where $P$ indicates the number of different classes of degrees ${\bf k}$) of the system of Eqs $(\ref{SpC})$ of elements
\bea
\hat{J}_{\alpha\beta\hat{\bf k};\gamma\alpha {\bf k}}=p_{\gamma\alpha}x_{\alpha\beta} \tilde{P}_{\beta\alpha}({\bf k}|\hat{\bf k})\left[{k}^{[\alpha,\gamma]}-\delta(\gamma,\beta)\right],
\label{J1}
\eea
where we have adopted the same notation as in Eq. $\ref{J1a}$ of the main text.
Above the  transition, this system must develop a set of non trivial solutions.
Therefore the transition point  is obtained by imposing that the 
 maximum eigenvalue $\hat{\Lambda}_J$ of the matrix $\hat{\bf J}$ satisfies
\bea
\hat{\Lambda}_J=1.
\eea


\section*{References}
\begin{thebibliography}{99}



\bibitem{Epidemics_review}
 Pastor-Satorras R,  Castellano C,  Van Mieghem P, and  Vespignani A 2015
{\it Rev.  Mod. Phys.} {\bf 87}  925 
\bibitem{Dynamics}
 Barrat A,  Barthelemy M, and  Vespignani A 2008 {\em Dynamical processes on complex networks} (Cambridge,Cambridge University Press)


\bibitem{Doro_crit} 
Dorogovtsev S N,  Goltsev A, and  Mendes J F F 2008 
{\it Rev. Mod. Phys.} {\bf 80} 1275 
\bibitem{Newman_book}
 Newman M E J 2010 {\em Networks: an introduction} (Oxford,Oxford University Press)
\bibitem{Laszlo}
 Barab\'asi A L 2016 {\em Network science} (Cambdrige,Cambridge University Press)

\bibitem{PhysReports}
 Boccaletti S,  \etal  M 2014 {\it Phys. Rep.} {\bf 544} 1 
 
\bibitem{Kivela}
 Kivel\"a M,  \etal 2014
{\it J. Complex Net.} {\bf 2} 203 
\bibitem{Arenas_review}
 De Domenico M,  Granell C,  Porter M A, and  Arenas A 2016
{\it Nature Physics} {\bf 12} 901 
\bibitem{diffusion}
{Gomez S,  \etal} 2013
{\it Phys. Rev. Lett.} {\bf 110}  {028701}.
\bibitem{Radicchi_diff}
 Radicchi F 2015 
{\it Nature Physics} {\bf 11} 597 
\bibitem{Boccaletti}
Cardillo A \etal  2013 {\it Sci. Rep.} {\bf 3} 1344 


\bibitem{Arenas_PNAS}
 De Domenico M, \etal 2014
{\it Proc.  Nat. Aca. Sci.} {\bf 111}  8351 
\bibitem{Arenas_awareness}
 Granell C,  G\'omez S, and  Arenas A 2013
{\it Phys. Rev. Lett.} {\bf 111}  128701 

\bibitem{Havlin1}
 Buldyrev S V,  Parshani R,  Paul G,  Stanley H E, and  Havlin S 2010
{\it Nature} {\bf  464} 1025  
\bibitem{Brummitt}
 Brummitt C D ,  d’Souza R M , and  Leicht E A 2012
{\it Proc.  Nat. Aca.  Sci.} {\bf 109} E680 


\bibitem{Boguna_epidemics}
{Saumell-Mendiola A,  Serrano M A, and Bogu\~n\'a M} 2012
{\it Phys. Rev. E} {\bf 86} {026106}
\bibitem{Cozzo}
 Cozzo E,   Banos R A,  Meloni S, and   Moreno Y 2013
{\it Phys. Rev. E} {\bf 88}  050801 


\bibitem{Yamir}
de Arruda G F,  Cozzo E,  Peixoto T P,  Rodrigues F A, and Moreno Y 2015
arXiv preprint arXiv:1509.07054 
\bibitem{Valdano}
 Valdano E, Ferreri L,  Poletto C and Colizza V 2015 {\it Phys. Rev. X} {\bf 5}, 021005 
\bibitem{Stanley_SIR}
Dickison M,  Havlin S, and  Stanley H E 2012
{\it Phys. Rev. E} {\bf 85} 066109 
\bibitem{Braunstein1}
 Buono C,  Alvarez-Zuzek L G,  Macri P A, and  Braunstein L A 2014
{\it PloS one} {\bf 9}  e92200 
\bibitem{Immunization}
 Zhao D,  Wang L,  Li S,  Wang Z,  Wang L, and  Gao B 2014
{\it PloS one} {\bf 9}  e112018 
\bibitem{Braunstein2}
 Buono C and  Braunstein L A 2015
{\it EPL (Europhysics Letters)} {\bf 109}  26001 
\bibitem{Yagan}
O. Yagan, and V. Gligor
{\it Phys. Rev. E} {\bf 86} (3), 036103 (2012).

\bibitem{Yagan2}
Y. Zhuang, A. Arenas, and O. Yagan arXiv preprint arXiv:1608.08237 (2016).

\bibitem{Yagan3}
Y. Zhuang,  and O. Yagan,
{\em IEEE Transactions on Network Science and Engineering} {\bf 3}, 211 (2016).


\bibitem{Newman_2002}
  Newman M E J 2002
{\it Phys. Rev. E} {\bf 66}  016128

\bibitem{Baxter}
 Baxter G J,  Dorogovtsev S N,  Goltsev A V, and  Mendes J F F 2012 
{\it Phys. Rev. Lett.} {\bf 109} 248701  



 
\bibitem{CDB_2016}
 Cellai D,  Dorogovtsev S N and  Bianconi G 2016
 {\it Phys. Rev. E} {\bf 94} 032301 
\bibitem{RB1}
 Bianconi  G and  Radicchi F 2016 arXiv preprint arXiv:1610.08708 
\bibitem{RB2}
 Radicchi F  and  Bianconi G 2016 arXiv preprint arXiv:1610.05378  

\bibitem{BD1}
 Bianconi G  and  Dorogovtsev SN 2014
{\it Phys. Rev. E} {\bf 89}  062814 
\bibitem{BD2}
 Bianconi G,   Dorogovtsev S N, and  Mendes J F F 2015
{\it Phys. Rev. E} {\bf 91}  012804 


\bibitem{dSouza}
 Leicht E A  and  d'Souza R M 2009
arXiv preprint arXiv:0907.0894 

\bibitem{Makse}
Reis S D S,  \etal  2014
{\it Nature Physics} {\bf 10} 762 



\bibitem{Cellai}
 Hackett A, Cellai D,  G\'omez S,  Arenas A, and  Gleeson J P 2016
{\it Phys. Rev. X} {\bf 6} 021002 

\bibitem{Lenka}
 Karrer B,  Newman M E J, and  Zdeborov\'a L 2014
{\it  Phys. Rev. Lett.} {\bf 113}  208702 
 
 \bibitem{Mezard}
Mezard M and  Montanari A 2009 
{\it Information, Physics and Computation} 
(Oxford University Press, Oxford)
\bibitem{Hamilton}
K. E. Hamilton, and L. P. Pryadko,
{\em Phys. Rev. Lett.} {\bf 113}: 208701 (2014).

 \bibitem{Radicchi_beyond}
 Radicchi F   and  Castellano C 2016
arXiv preprint arXiv:1602.07140 

\bibitem{Newman_dynamics}
B. Karrer,  and M. E. J. Newman,
{\it Physical Review E} {\bf 82}, 016101 (2010).
\bibitem{DallAsta}
F. Altarelli, A. Braunstein, L. Dall’Asta, J. R. Wakeling, and R. Zecchina, 
{\em Phys. Rev. X} {\bf 4},  021024 (2014).
\bibitem{Saad}
A. Y. Lokhov,  and David Saad,
arXiv preprint arXiv:1608.08278 (2016).




\end{thebibliography}
\end{document}